\documentclass[twocolumn,superscriptaddress]{aastex63}

\accepted{to PSJ Feb 11, 2024}

\usepackage{graphicx}
\usepackage{bm}
\usepackage{mhchem}
\usepackage{longtable}
\usepackage{supertabular}
\usepackage{textpos}
\usepackage{amsmath}

\usepackage{wasysym}

\usepackage{soul}

\newcommand{\beginsupplement}{%
        \setcounter{table}{0}
        \renewcommand{\thetable}{S\arabic{table}}%
        \setcounter{figure}{0}
        \renewcommand{\thefigure}{S\arabic{figure}}%
        \setcounter{equation}{0}
        \renewcommand{\theequation}{S\arabic{equation}}%
     }

\shorttitle{Impact experiments on organic haze particles}
\shortauthors{Pearce et al.}

\begin{document}

\title{Towards Prebiotic Chemistry on Titan: Impact experiments on organic haze particles}

\author{Ben K. D. Pearce*}
\affiliation{Department of Earth and Planetary Science, Johns Hopkins University, Baltimore, MD, 21218, USA}
\thanks{bpearce6@jhu.edu; chaohe23@ustc.edu.cn}

\author{Sarah M. H{\"o}rst}
\affiliation{Department of Earth and Planetary Science, Johns Hopkins University, Baltimore, MD, 21218, USA}

\author{Christopher J. Cline}
\affiliation{NASA Johnson Space Center, Astromaterials Research and Exploration Science, Mail Code X13, 2101 NASA Parkway,Houston, Texas 77058, USA}

\author{Mark J. Cintala}
\affiliation{NASA Johnson Space Center, Astromaterials Research and Exploration Science, Mail Code X13, 2101 NASA Parkway,Houston, Texas 77058, USA}

\author{Chao He*}
\affiliation{School of Earth and Space Sciences, University of Science and Technology of China, Hefei, China}
\affiliation{Department of Earth and Planetary Science, Johns Hopkins University, Baltimore, MD, 21218, USA}

\author{Joshua A. Sebree}
\affiliation{Department of Chemistry and Biochemistry, University of Northern Iowa, Cedar Falls, IA, USA}

\author{Shannon M. MacKenzie}
\affiliation{Applied Physics Laboratory, Johns Hopkins University, Space Exploration Sector, Laurel, MD 20723, USA}

\author{R. Terik Daly}
\affiliation{Applied Physics Laboratory, Johns Hopkins University, Space Exploration Sector, Laurel, MD 20723, USA}

\author{Alexandra J. Pontefract}
\affiliation{Applied Physics Laboratory, Johns Hopkins University, Space Exploration Sector, Laurel, MD 20723, USA}

\author{Cara Pesciotta}
\affiliation{Department of Earth and Planetary Science, Johns Hopkins University, Baltimore, MD, 21218, USA}

\begin{abstract}
{\bf 
Impacts are critical to producing the aqueous environments necessary to stimulate prebiotic chemistry on Titan's surface. Furthermore, organic hazes resting on the surface are a likely feedstock of biomolecules. In this work, we conduct impact experiments on laboratory-produced organic haze particles and haze/sand mixtures and analyze these samples for life's building blocks. Samples of unshocked haze and sand particles are also analyzed to determine the change in biomolecule concentrations and distributions from shocking. Across all samples, we detect seven nucleobases, nine proteinogenic amino acids, and five other biomolecules (e.g., urea) using a blank subtraction procedure to eliminate signals due to contamination. We find that shock pressures of 13 GPa variably degrade nucleobases, amino acids, and a few other organics in haze particles and haze/sand mixtures; however, certain individual biomolecules become enriched or are even produced from these events. Xanthine, threonine, and aspartic acid are enriched or produced in impact experiments containing sand, suggesting these minerals may catalyze the production of these biomolecules. On the other hand, thymine and isoleucine/norleucine are enriched or produced in haze samples containing no sand, suggesting catalytic grains are not necessary for all impact shock syntheses. Uracil, glycine, proline, cysteine, and tyrosine are the most unstable to impact-related processing. These experiments suggest that impacts alter biomolecule distributions on Titan's surface, and that organic hazes co-occurring with fine-grained material on the surface may provide an initial source for further prebiotic chemistry on Titan.
}

\end{abstract} 

\keywords{Titan --- impacts --- organic hazes --- nucleobases --- amino acids --- astrobiology}

\section*{Introduction}

The atmosphere of Saturn's moon Titan contains multiple layers of organic haze \citep{Atreya2007,2017JGRE..122..432H}. These solid particles are naturally produced in \ce{N2}/\ce{CH4}-rich atmospheres via chemical reactions initiated by collisions, ultraviolet (UV) radiation, or thermal ionization and dissociation \citep{Trainer_et_al2006, 2017JGRE..122..432H}. Laboratory analogs of these haze particles contain a myriad of biomolecules including the nucleobases of RNA and DNA, as well as amino acids and their intermediates \citep{Pearce_et_al2023,Sebree_et_al2018}. 

Titan's crust is predominantly water ice, but the surface is also covered in a variety of organic sediments believed to be sourced from the atmospheric hazes \citep{2017JGRE..122..432H}. Impacts from comets and planetesimals on the icy surface produce melt pools that can survive up to tens of thousands of years \citep{Reference441}. These post-impact environments are key sites for potential complex prebiotic chemistry. Crater counts on Saturnian moons suggest hundreds to thousands of impacts producing craters $>$10 km in diameter have occurred on Titan \citep{Zahnle_et_al2003}. However, weathering processes such as aeolian infilling and fluvial erosion actively modify Titanian craters, leaving behind only a few dozen craters that have been observed by Cassini RADAR \citep{Hedgepeth_et_al2020}. The Dragonfly mission to Titan will sample impacted materials surrounding Selk crater in the mid-2030's, which will provide some ground truth on the extent of the prebiotic chemistry that occurs on Titan \citep{Barnes_et_al2021}. In the meantime, impact simulations on Titan haze particle analogs in the lab can predict what Dragonfly might find.

Most of Titan’s observed impact craters are located in the equatorial region where the surface is dominated by organic sands \citep{2017JGRE..122..432H}. These sand grains are larger than typical haze particles in Titan's atmosphere. For this reason, it is unclear whether these sands are composed of haze material, and, if they are, how the material processed into larger particles \citep{2017JGRE..122..432H}. As the bulk composition of the sand on Titan is unknown, it is unclear what fraction of surface haze particles would be reactive in the impact conditions. For this reason, we mix our experimental organic haze particles with a quartz sand commonly used in impact studies \citep{Cline_Cintala2022} in order to simulate the physical effects of impact conditions containing both large and small grains.

It is currently unknown how the biomolecule content of organic hazes is modified by the initial shock from impact events. Previous shock experiments investigating impacts on icy planetary or cometary surfaces led to the production of amino acids such as glycine, alanine, $\alpha$-aminoisobutyric acid, and isovaline \citep{2013NatGe...6.1045M}. On the other hand, shock experiments on samples of amino acids and peptides in artificial meteorites, and aqueous solutions of amino acids, reveal that these molecules can also be destroyed during the chaos of impact, with survival percentages dependent on shock pressure \citep{Bertrand_et_al2009,Blank_et_al_2001}. \citet{Bertrand_et_al2009} found that amino acids and dipeptides generally had an $>$30\% survival percentage for a shock pressure of 12 GPa, whereas the survival percentage dropped to 0--4\% for 28.9 GPa. Similarly, \citet{Blank_et_al_2001} found that a significant fraction ($\sim$36--72\%) of amino acids in aqueous solution survived shock pressures ranging from 5--21 GPa. Finally, experiments simulating the thermal alteration of organics during hypervelocity capture suggest that there may be only slight thermal decomposition of organic samples for these short (e.g., 10 ms) heating events \citep{Bowden_et_al2008}.

In this study, we performed impact experiments on simulated Titan organic haze particles and analyzed the shocked material for biomolecule composition using gas chromatography/mass spectrometry/mass spectrometry (GC/MS/MS) methods. Our experimental shock pressures of $\sim$13 GPa are roughly matching or slightly below the average impact velocity (10 km/s) on Titan's icy surface \citep{Zahnle_et_al2003} as suggested by shock simulations onto H2O ice \citep{Kraus_et_al2011}. Hydrocode models of planetary surfaces suggest this impact energy will completely melt but not completely vaporize the impact region \citep{Pierazzo_et_al1997}.





\section*{Methods}

\subsection*{Haze Production}

Titan haze particle analogs are produced using the Planetary Haze Research (PHAZER) experimental setup \citep{He_et_al2017}. This setup contains a vacuum flow system and stainless steel chamber within which an energy source can be applied to simulate ultraviolet, cosmic ray, or lightning chemistry in planetary atmospheres. The energy source we used for this work was a cold plasma discharge, produced by applying a voltage differential of 6000 V between two electrodes. The electrons produced in the plasma collide with the gas mixture, breaking chemical bonds in a similar fashion to short-wave upper atmospheric UV light ($\lesssim$ 110nm) \citep{Pearce_et_al2022b,Cable_et_al2012} and energetic particles \citep{Pearce2020a}. This bond-breaking initiates the chemical pathways that lead to the production of organic haze particles. For a schematic of the PHAZER setup, see \citet{Pearce_et_al2022b}.

Our experimental protocol matches that of previous PHAZER studies \citep{Pearce_et_al2023,Pearce_et_al2022b,Moran_et_al2020,He_et_al2020b,He_et_al2020a,He_et_al2019,Horst_et_al2018a,He_et_al2018b,He_et_al2018a}.
To summarize, we prepared the gas mixtures using high-purity gases (\ce{N2}-99.9997\%, \ce{CH4}-99.999\%, and \ce{CO}-99.99\%; Airgas). The gases flowed through a cooling coil submerged in liquid nitrogen prior to reaching the reaction chamber to obtain 90$\pm$5 K atmospheric temperatures \citep{He_et_al2017}. We applied a gas flow of 10 standard cubic centimeters per minute and initiated the 170 W/m$^2$ cold plasma source, which produces electrons and ions in the 5--15 eV range. The chamber pressure for this experiment was 1.95 Torr.

The organic haze particles were produced using a gas mixture representing Titan's atmosphere of {\ce{N2}}:{\ce{CH4}}:{\ce{CO}} = 94.8:5:0.2. Experiments were run for approximately 3 days until an orange-brown powder formed on the walls of the chamber, as seen through the chamber window. We then turned off the plasma source and gas flow, sealed the chamber, and transferred it into a dry ($<$0.1 ppm \ce{H2O}), oxygen-free ($<$0.1 ppm \ce{O2}) \ce{N2} glovebox (I-lab 2GB; Inert Technology) where samples of the orange-brown solid haze particles were collected with sterilized scoopulas from the walls and base of the chamber and stored in small glass vials. Samples were fabricated by mixing the simulated organic haze with powdered novaculite (fine-grained quartzite sourced from Arkansas, USA) in glass vials to yield three different sample compositions of 100\%, 50\%, and 10\% organic haze by mass. Samples were then sealed with parafilm, safely wrapped and secured in an envelope, and express shipped to NASA Johnson Space Center for the shock experiments.

\subsection*{Shock Experiments}

\begin{figure}[!hbtp]
\centering
\includegraphics[width=\linewidth]{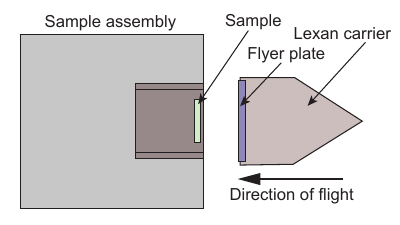}
\caption{Schematic of sample assembly used in the flat-plate accelerator experiments. Not drawn to scale. \label{Shock_Schematic}}
\end{figure}

Shock-recovery experiments were performed using the single-stage, gunpowder-driven, flat-plate accelerator that is housed in the Experimental Impact laboratory at NASA Johnson Space Center. Details regarding the shock-reverberation technique and general hardware used for these experiments can be found in \citet{Horz1970} and \citet{Gibbons_et_al1975}. Samples were unpacked, and 100 mg of each mixture was then loosely packed into a tungsten alloy (HD-17, 90\% W, 6\% Ni, and 4\% Cu) inner sample container that was subsequently pressed into a larger stainless-steel outer assembly and stored under an atmosphere of 50 mtorr until the time of the experiment (Fig.~\ref{Shock_Schematic}). 

For each shock experiment, we placed the assembly inside the impact chamber, evacuated the atmosphere to ~40 mtorr, and impacted the assembly with an aluminum flyer plate at $\sim$900 m s$^{-1}$. Peak pressure for each shot was determined using a one-dimensional impendence-matching technique and the Hugoniots for the metals of the flyer plate and sample assemblies. The achieved reverberation pressures for the 100\%, 50\%, and 10\% organic haze samples were 12.9, 12.5, and 13.2 GPa, respectively. The duration of the shocked state for experiments at this scale is from 0.1--1 $\mu$s. Photos of the flyer plate immediately before impact were used to determine the tilt angle of the projectile, and all experiments showed maximum tilts of $<$ 2.7$^{\circ}$. After recovering each shocked assembly, the bulk of the metal was machined away to within a few millimeters of the shocked sample (performed slowly so as not to raise the temperature of the sample). The sample assemblies were then wrapped and secured in a box, and delivered to Johns Hopkins University. The remaining metal was then manually pried off and the sample removed and stored in glass vials with parafilm. The samples were only exposed to atmosphere briefly (a few minutes) when we opened the assembly and collected the sample. These glass vials were then wrapped and secured in an envelope, and brought to the University of Northern Iowa for analysis.

Notably, for all three samples containing haze particles, a viscous substance formed on the outside of the sample assembly that may have escaped through one or more cracks from the sample interior. An initial GC/MS/MS analysis was performed on the viscous substance---and it was tentatively found to contain biomolecules---however, given its limited quantity, we ran out of the viscous substance before we developed our most accurate blank subtraction GC/MS/MS protocol. Given the large uncertainties in our initial analysis, the biomolecule content of the viscous substance was not included in this work.

\subsection*{Biomolecule Analysis}

Analysis of all samples (haze particles, sand, and mixtures) was performed using GC/MS/MS protocols originally developed in \citet{Sebree_et_al2018} and updated in \citet{Pearce_et_al2023}. GC/MS/MS is a highly selective and sensitive analysis technique that makes use of all three quadrupoles of the Agilent 7000c triple quadrupole GC/MS in order to concentrate and isolate specific molecules of interest. This method differs from extracted ion chromatogram GC/MS analysis, where quantitation ions are extracted from a total ion chromatogram. Instead, the GC/MS/MS technique only searches for one or two species within exclusive ranges of retention times referred to as gates. Within each gate, a selected parent ion is trapped and pre-concentrated at the first quadrupole. This is followed by neutral gas collision and fragmentation at the second quadrupole, where finally, the third quadrupole allows only the qualifier and quantifier daughter ions associated with the fragmentation to pass through the detector. This procedure is well suited for analyzing complex mixtures requiring high sensitivity such as those in this study. We note that with this technique the mass spectra for each gate generally only contain one qualifier and one quantifier daughter peak. The only exception to this is when two species are in the same gate due to their similar retention times. In this case, the two quantifiers from the mass spectrum can be used to deduce the relative fraction of the GC peak that belongs to each species (e.g., uracil and proline). Analysis for this work was done in parallel with the early Earth haze analysis performed in \citet{Pearce_et_al2023}; thus, the standards used to develop calibration curves are are the same as in \citet{Pearce_et_al2023}.

Two standards were analyzed for this work: an industrial-made physiological amino acid standard (Sigma Aldrich) with 27 detectable components at 0.25 $\mu$mol mL$^{-1}$ biomolecule$^{-1}$, and a custom made standard containing seven nucleobases (Sigma Aldrich, all 99.0\%). To make the latter standard, we dissolved seven nucleobases in a 0.1 M solution of NaOH at a concentration of 0.5 mg mL$^{-1}$ nucleobase$^{-1}$. To measure a range of standard concentrations for calibration curves and limits of detection/quantification, we made separate dilutions of these standards of $\frac{5}{100000}$, $\frac{20}{100000}$, and $\frac{200}{100000}$ to use as base concentrations from which we pulled multiple volumes for GC/MS/MS analysis.

The procedure for standard preparation was as follows: first, we pipetted multiple volumes ranging from 5--50$\mu$L of each standard dilution into to separate GC vials and dried the vials in the oven at 40$^{\circ}$C under \ce{N2} flow. After the solutions were dry, we added 100 $\mu$L of dichloromethane (\ce{CH2Cl2}) and heated at 40$^{\circ}$C under \ce{N2} flow to evaporate any remaining solvent. Next, we added 30$\mu$L of dimethylformamide (DMF) and 30 $\mu$L of N-tert-butyldimethylsilyl-N-methyltrifluoroacetamide (MTBSTFA) to the vials and allowed the solutions to derivatize for 30 minutes under \ce{N2} flow at 80$^{\circ}$C. Lastly, we diluted samples with 100$\mu$L of \ce{CH2Cl2}, and injected volumes of 0.5, 0.8, 1, and/or 2$\mu$L into the GC triple quad running in MS/MS mode for analysis. Samples took approximately 25 minutes to run. Two procedural blanks were run through the GC/MS/MS prior to each standard run, and the chromatogram of the second blank was subtracted from the standard chromatogram for analysis.

Compound separation was performed using a Restek capillary column (RTX-5MS) with helium flow held at 1.3 mL min$^{-1}$. The initial operating temperature was 100$^{\circ}$C, with a ramp up rate of 10$^{\circ}$C min$^{-1}$ to a final temperature of 270$^{\circ}$C that was held for 11.5 minutes. We used collisionally induced dissociation energies in the range of 10--50 eV for GC/MS/MS analysis. The most effective collision energies for the species in each gate were determined in \citet{Sebree_et_al2018} by running multiple energies in MS mode and seeing which resulted in the highest daughter mass spectrum peaks. See Table~\ref{RetentionTimes} for details on GC gate retention times, ion peaks, and collision energies.

In Figures~\ref{NucleobaseCalCurve} and \ref{PhysioCalCurve}, we display the calibration curves for nucleobases and amino acids/other biomolecules, respectively. These curves were fit to 3--5 data points that a) covered the range of peak areas measured in our samples, b) provided the lowest $\chi^2$ value for a least squares linear fit, and c) had data points above the limit of quantification. The one exception was urea, for which some samples contained factors of 1.3--5 times greater peak areas than our most concentrated standard. In these cases, we calculated urea concentrations via extrapolation of the calibration curve, and suggested the values as upper limits. We placed the peak areas measured in the unshocked haze particle samples on the calibration curves to demonstrate quantification. See the supplement of \citet{Pearce_et_al2023} for further details on the standards and Table~\ref{LODLOQTable} for further analysis on limits of detection and quantification.

Peak areas and uncertainties were calculated using an interactive Python program developed in \citet{Pearce_et_al2023}. The program automatically detects peaks within user-defined gates, and requires the user to select the noise regions in each gate in order to calculate the average noise floor for the area calculation. Uncertainties in the area are automatically calculated by varying the noise floor by $\pm$ one standard deviation. However, uncertainties in the calculated concentrations within our organic haze particle samples take into account both the uncertainty in the GC peak area, as well as the uncertainty in the calibration curve line of best fit. The visual aspect of the program allows the user to validate whether peaks are at the correct retention times based on the standards. Furthermore, the program determines signal-to-noise ratios, allowing the user to screen out signals below SNR = 3. The program is open source and can be downloaded from Zenodo \citep{Pearce_Chromatogram_2023}.

Samples were prepared for analysis using the protocols from \citet{Pearce_et_al2023}. Haze, sand, and sand/haze samples were weighed and dissolved in a 5.0 mg mL$^{-1}$ 50:50 solution of methanol and acetone. Vials were agitated and then centrifuged at 10,000 rpm for 10 minutes. To encourage further organic dissolution and separation, vials were rotated 180 degrees and centrifuged again for an additional 10 minutes. Then, we used a Pasteur pipette to transfer the supernatants to GC vials. Next, we dried and derivatized the solutions using the same methods described for the standards above. We injected 1$\mu$L of the solutions into the GC triple quad running in MS/MS mode for analysis. Two blank solutions of 50:50 methanol/acetone were carried through the same procedure, and ran through the GC/MS prior to each sample run. We subtracted the blank chromatogram that was run directly prior to each sample chromatogram for analysis. This procedure was developed in this way due to a comprehensive noise analysis performed in \citet{Pearce_et_al2023}, which revealed that underivatized biomolecules from the standards can get trapped in the GC inlet and become derivatized and released into the GC column upon subsequent blank or sample runs. Trapping cannot be completely diminished due to the $<$ 100\% derivatization efficiency of biomolecule samples, as well as the derivatization decay that occurs while the samples are transported to the GC/MS or waiting in the rotating sample queue for GC/MS/MS analysis. 

In \citet{Pearce_et_al2023}, we performed two contaminant analyses to test whether this trapping occurs. First, after a standard was run through the GC-MS, we ran 17 consecutive blanks containing derivatizing agents. From this test, we found that the standard peaks decreased with each additional run, indicating that this procedure slowly washes/removes underivatized biomolecules out of/from the inlet. In no case did a subsequent wash run produce higher peaks than the previous run. Second, we ran blanks without derivatizer and found no contaminant peaks, suggesting that underivatized biomolecules are derivatized and able to vaporize out of the inlet when derivatizer was present in the blank run. Because we cannot completely remove the contaminant peaks from procedural blank runs containing derivatizer, two blanks are run before each sample to wash the inlet, and then the second blank chromatogram was subtracted from the sample chromatogram prior to analysis.

\section*{Results}

\begin{figure*}[!hbtp]
\centering
\includegraphics[width=\linewidth]{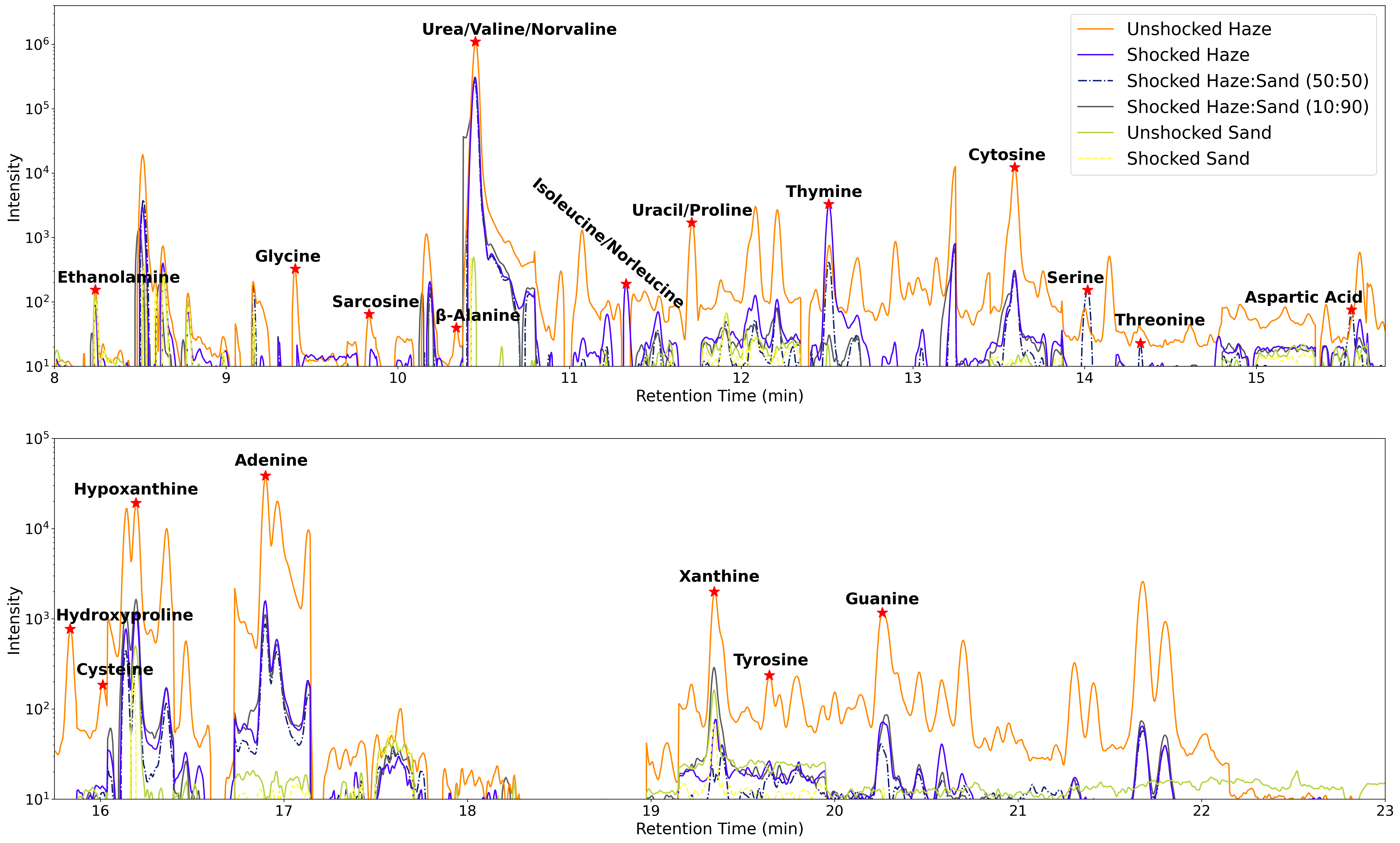}
\caption{Blank-subtracted GC/MS/MS chromatograms for the six shocked and unshocked samples in this study. \label{HazeSpectra}}
\end{figure*}

In Figure \ref{HazeSpectra}, we display the GC/MS/MS chromatograms for the six samples in this study. We shocked three haze samples, containing A) 100\% haze particles, B) 50:50 haze:sand, and C) 10:90 haze:sand. We also analyzed unshocked haze particles, as well as unshocked and shocked sand samples. We detected seven nucleobases, nine proteinogenic amino acids, and five other organics (e.g., urea) across all samples. We summarize the detected biomolecules and their concentrations in Table~\ref{ConcentrationTable}.


\begin{table*}[ht!]
\centering
\caption{Measured mass fractions of nucleobases, amino acids, and other organics in shock experiments. Uncertainties are the combination of GC peak area uncertainties, and the calibration curve uncertainties. A few values are upper bounds as their GC peak measurements fall between the limits of detection (LOD) and the limits of quantification (LOQ). See Table~\ref{LODLOQTable} for more detail. Units are $\mu$g biomolecule / g sample, or ppm. \label{ConcentrationTable}} 
\begin{tabular}{lcccccc}
\\
\multicolumn{1}{l}{Biomolecule} &  
\multicolumn{1}{l}{Unshock. Haze} & 
\multicolumn{1}{l}{Shock. Haze} &
\multicolumn{1}{l}{Shock. H:S 50:50} & 
\multicolumn{1}{l}{Shock. H:S 10:90} &
\multicolumn{1}{l}{UnShock. Sand} &
\multicolumn{1}{l}{Shock. Sand} 
\\[+2mm] \hline \\[-2mm]
{\bf Nucleobases} & & & \\
Adenine & 127$\substack{+17 \\ -2}$ & 7.5$\substack{+2.0 \\ -0.1}$ & 6.7$\substack{+2.7 \\ -0.2}$ & 5.6$\substack{+2.5 \\ -0.3}$ & - & - \\
Cytosine & 58$\substack{+2 \\ -4}$ & 2.7$\substack{+0.4 \\ -0.1}$ & 3.7$\substack{+0.1 \\ -0.1}$ & 2.6$\substack{+0.3 \\ -0.1}$ & - & - \\
Guanine & 44.8$\substack{+0.4 \\ -2.3}$ & 19$\substack{+4 \\ -1}$ & 17$\substack{+3 \\ -2}$ & 20$\substack{+4 \\ -1}$ & - & -   \\
Hypoxanthine & 38.6$\substack{+0.6 \\ -0.1}$ & 27.1$\substack{+1.5 \\ -0.1}$ & 19.8$\substack{+0.2 \\ -0.2}$ & 33.8$\substack{+1.3 \\ -0.1}$ & 18.3$\substack{+1.1 \\ -0.1}$ & 16.3$\substack{+0.2 \\ -1.3}$\\
Xanthine & 30.8$\substack{+4.7 \\ -0.4}$ & 11.3$\substack{+3.8 \\ -0.6}$ & - & 51$\substack{+3 \\ -1}$ & 26.1$\substack{+3.5 \\ -0.1}$ & 17$\substack{+6 \\ -1}$ \\
Uracil & 5.0$\substack{+0.1 \\ -1.2}$ & - & - & - & - & - \\
Thymine & 2.1$\substack{+0.0 \\ -0.5}$ &2.9$\substack{+0.0 \\ -0.3}$ & 1.84$\substack{+0.01 \\ -0.32}$ & - & 1.155$\substack{+0.001 \\ -0.001}$ & - \\
{\bf Protein. Amino A.} & & & \\
Tyrosine & 91$\substack{+31 \\ -15}$ & - & - & - & - & - \\
Valine/Norvaline & 80$\substack{+0 \\ -0}$ & 24.5$\substack{+0.0 \\ -0.6}$ & 47$\substack{+0 \\ -2}$ & 25.9$\substack{+0.0 \\ -0.8}$ & - & - \\
Cysteine & $<$11.1$\substack{+0.5 \\ -0.5}$ & - & - & - & - & - \\
Proline & 1.80$\substack{+0.03 \\ -0.26}$ & - & - & - & - & - \\
Glycine & 0.78$\substack{+0.21 \\ -0.01}$ & - & - & - & - & - \\
Isoleucine/Norleucine & - & $<$0.35$\substack{+0.08 \\ -0.07}$ & - & - & - & - \\
Aspartic Acid & - & - & 8$\substack{+1 \\ -1}$ & - & - & - \\
Threonine & - & - & 7.6$\substack{+0.0 \\ -0.4}$ & - & - & -\\
Serine & 6.5$\substack{+0.1 \\ -0.1}$ & - & 7.4$\substack{+0.1 \\ -1.0}$ & - & - & - \\
{\bf Other Biomolecules} & & & \\
Urea & 4692$^a$$\substack{+2 \\ -2}$ & 791$^a$$\substack{+0 \\ -0}$ & 948$^a$$\substack{+0 \\ -0}$ & 977$^a$$\substack{+1 \\ -1}$ & 6.24$\substack{+0.01 \\ -0.01}$  & -\\
Sarcosine & $<$12.1$\substack{+0.3 \\ -0.3}$ & - & - & - & - & - \\
Hydroxyproline & 9.4$\substack{+2.3 \\ -1.2}$ & - & - & - & -  & - \\
$\beta$-Alanine & $<$0.53$\substack{+0.07 \\ -0.07}$ & - & - & - & - & - \\
Ethanolamine & 0.44$\substack{+0.03 \\ -0.02}$ & 0.18$\substack{+0.01 \\ -0.01}$ & 0.37$\substack{+0.01 \\ -0.01}$ & $<$0.06$\substack{+0.01 \\ -0.01}$ & 0.27$\substack{+0.02 \\ -0.01}$ & $<$0.30$\substack{+0.02 \\ -0.02}$\\
\\[-2mm] \hline
\multicolumn{7}{l}{\footnotesize $^a$ Upper limit based on upper bound extrapolation of calibration curve.} \\
\end{tabular}
\end{table*}


Unexpectedly, we detected a sulfur-containing amino acid, cysteine, in the unshocked haze sample right at the limit of detection (SNR = 3.3). We did not purposefully introduce sulfur into the gas mixture for the Titan haze production experiments. It is possible that this small peak is a non-sulfur containing molecule with similar parent and daughter ions as cysteine. However, it is also possible that cysteine was produced in our atmospheric haze experiments from small sulfur impurities in our gas mixtures. It is unlikely that this peak is contaminant introduced during sample preparation or by the GC inlet, as both blank chromatograms run prior to this sample showed no peak at the cysteine retention time.

\begin{figure}[!hbtp]
\centering
\includegraphics[width=\linewidth]{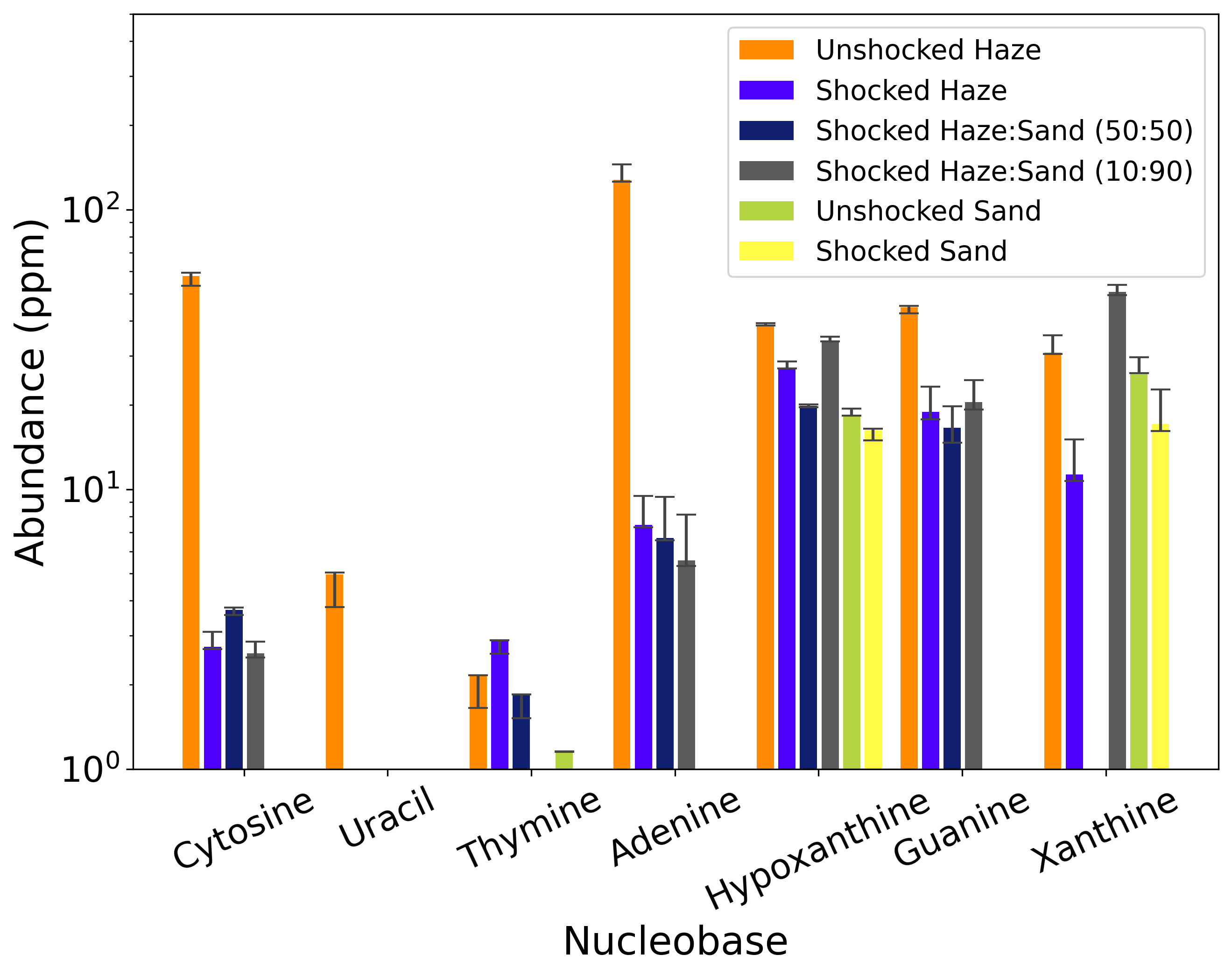}
\caption{Nucleobase abundances measured in the six shocked and unshocked samples in this study. Nucleobase order is from lowest to highest molar mass. Uncertainties are the combination of GC peak area uncertainties, and the calibration curve uncertainties. Comparisons can be made between any two samples, as units are $\mu$g biomolecule / g sample (e.g., haze, sand, or haze/sand), or ppm. \label{Nucleobases}}
\end{figure}

In Figure~\ref{Nucleobases}, we display the nucleobase mass abundances with respect to the sample mass (i.e., haze particles, sand particles, or haze/sand mixture). The unshocked haze particles have the highest summed nucleobase abundances. Individual mass fractions range from 2--127 ppm by mass. Shocking haze particles reduces the total nucleobase concentrations within them by a factor of $\sim$3--6. However, we see slight increases of individual nucleobase concentrations such as thymine and xanthine from shocking two samples: thymine by a factor of 1.4 in the 100\% shocked haze sample, and xanthine by a factor of 1.7 in the 10:90 shocked haze:sand sample (compared to the unshocked haze). Uracil is the most unstable nucleobase to impacts, as it was undetected in any shocked sample.

Unshocked quartz sand contains thymine, xanthine, and hypoxanthine concentrations at factors of 1.2--2.1 lower than the unshocked haze sample. However, we only detected xanthine and hypoxanthine in the shocked sand sample, and they were depleted in concentration by a factor of 1.1--1.5 with respect to the unshocked sand sample. For future shock experiments of haze particles, it will be beneficial to sterilize the sand prior to mixing it with the haze samples.

Lastly, although we do not include a biomolecule analysis of the viscous substance that formed on the outside of shock assembly for the samples containing organic haze particles, we weighed the viscous substance and found it to be $\sim$2\% the mass of the shocked haze. We suggest that the viscous substance is likely an escaped byproduct of haze material interacting with substances found inside the impact chamber during a shot, such as combusted gun powder and molten lexan.

\begin{figure}[!hbtp]
\centering
\includegraphics[width=\linewidth]{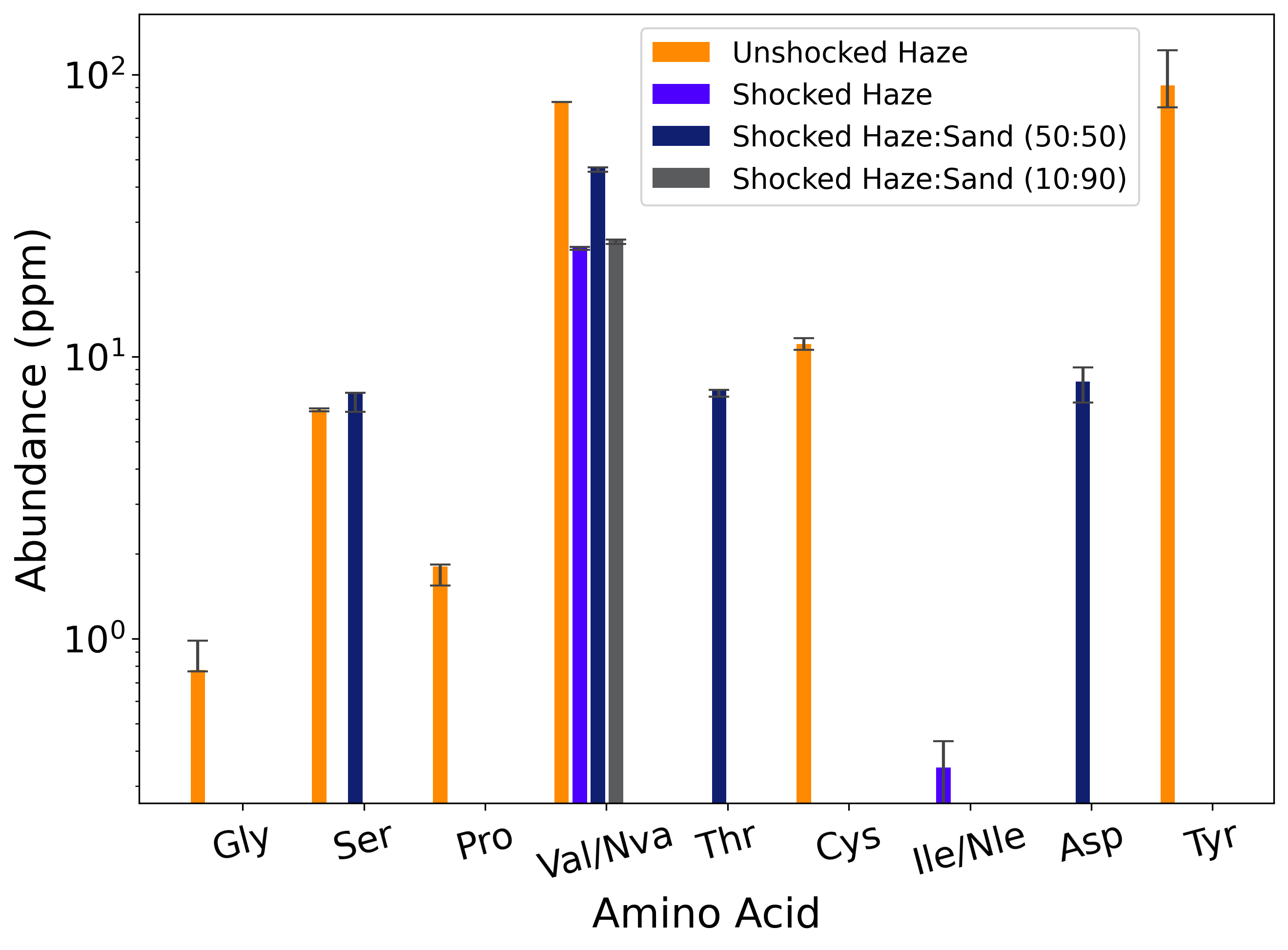}
\caption{Proteinogenic amino acid abundances measured in the shocked and unshocked samples in this study. Amino acid order is from lowest to highest molar mass. Uncertainties are the combination of GC peak area uncertainties, and the calibration curve uncertainties. Comparisons can be made between any two samples, as units are $\mu$g biomolecule / g sample (i.e., haze, sand, or haze/sand), or ppm. \label{Aminos}}
\end{figure}

In Figure~\ref{Aminos}, we display the proteinogenic amino acid abundances in the unshocked and shocked samples in this study. Abundances range from 0.35--91 ppm by mass. Shocking haze samples also reduces total amino acid content, by a factor of 3--8. Glycine, proline, cysteine, and tyrosine are the most unstable to shocks, as we did not detect them in any of the shocked haze samples. On the other hand, we detected isoleucine/norleucine, aspartic acid, and threonine in the 100\% haze or 50:50 shocked haze:sand samples, but not in the unshocked haze sample. This suggests that these three amino acids are products from shocking organic haze particles.

\begin{figure}[!hbtp]
\centering
\includegraphics[width=\linewidth]{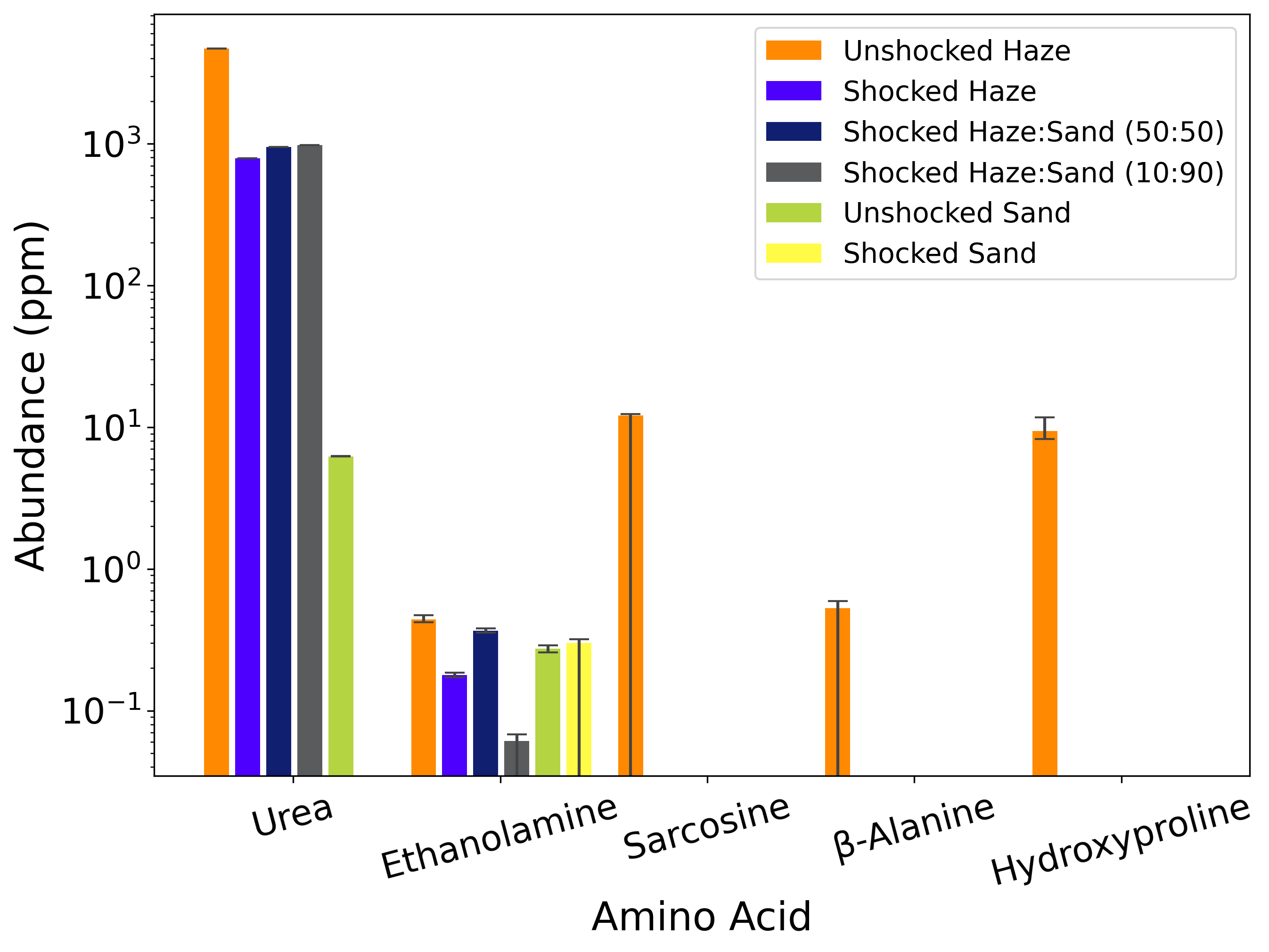}
\caption{Other biomolecule abundances measured in the shocked and unshocked samples in this study. Amino acid order is from lowest to highest molar mass. Uncertainties are the combination of GC peak area uncertainties, and the calibration curve uncertainties. Comparisons can be made between any two samples, as units are $\mu$g biomolecule / g sample (i.e., haze, sand, or haze/sand), or ppm. \label{Otherbios}}
\end{figure}

In Figure~\ref{Otherbios}, we display the abundances of non-proteinogenic amino acids, as well as amino acid derivatives and metabolites in the six samples in this study. The unshocked haze sample has the highest abundances of these organics, ranging from 0.44 to $<$ 4692 ppm by mass. Similar to nucleobases and amino acids, shocking the haze samples reduces the total content of these biomolecules by a factor of $\sim$5--6. Sarcosine, $\beta$-alanine, and hydroxyproline are the least stable to impacts, as we did not detect them in any of the shocked samples.

\section*{Discussion}

Our experiments suggest that the biomolecules anticipated to be in Titan's organic hazes are generally reduced in concentration when subjected to shock pressures of $\sim$13 GPa. On the other hand, some individual nucleobases and amino acids are enriched or produced from the high temperature and high pressure conditions of our impact experiments. 

\citet{2013NatGe...6.1045M} performed shock experiments on ice mixtures analagous to those found in a comet (e.g., \ce{NH4OH}:\ce{CO2}:\ce{CH4OH}) and found several amino acids were produced, including glycine, alanine, and norvaline. These results contrast ours, as we only detected norvaline out of these three species in any of our shocked organic haze experiments; and in lower concentrations than measured in the unshocked haze particles.

Impact experiments on amino acids in saponite clay showed that norvaline, glycine, aspartic acid, and serine were fairly stable to 12--15 GPa shocks \citep{Bertrand_et_al2009}. Similarly, impact experiments on amino acids in solution showed norvaline to be one of the most stable species, and proline to be fairly stable \citep{Blank_et_al_2001}. This is fairly consistent with our results, as we detected norvaline in all of our shocked haze samples, as well as aspartic acid and serine in at least one of our shocked haze samples. However, we did not detect glycine or proline in any of our shocked haze samples. Overall, we do expect there to be discrepancies between impact experiments performed at different pressures and scales. For example, impact experiments by \citet{Bowden_et_al2009} found that concentrations in impact ejecta vary with respect to sample location.



In recent biomolecule analyses of organic hazes heated to 200 $^{\circ}$C, uracil was the only nucleobase that survived (albeit reduced by a factor of two) of the seven studied nucleobases after 7 days \citep{Pearce_et_al2023}. This contrasts our results, which show that uracil is the most unstable nucleobase in regards to withstanding shocks. This discrepancy suggests that thermal degradation at 200 $^{\circ}$C and shock degradation affect nucleobases differently. We did not use a thermocouple to measure the peak temperatures from our shock experiments, and given the shock wave bounces between the metal that sandwiches the sample, it is too difficult to provide an accurate temperature estimate; however, if it is greater than 200 $^{\circ}$C, this could be the reason for the differences in the degradation of nucleobases between these two experiments.

There were no increases to nucleobase or amino acid concentrations in the haze particle heating experiments in \citet{Pearce_et_al2023}. However, in this study, thymine and isoleucine/norleucine were enriched/produced in the 100\% haze shock experiment, and xanthine, threonine, and aspartic acid were enriched/produced in haze:sand shock experiments. This suggests that the high-pressure, high-temperature environment created by impacts, as well as the potentially catalytic behavior of quartz sand provides distinct conditions for the production and modification of biomolecules in organic hazes.

Quartz sand may have catalytic effects on the chemistry of biomolecules in organic hazes when subjected to impacts. This is most obvious in the 50:50 haze:sand shock experiments, which produced threonine and aspartic acid. This could be due to mineral adsorption, which is known to catalyze a variety of organic reactions \citep{Hashizume2015} and stabilize amino acids and nucleobases against degradation \citep{Poggiali_et_al2020}. Experiments by \citet{Hashizume_et_al2010} showed that absorption of nucleobases by montmorillonite decreases in the order adenine $>$ cytosine $>$ uracil. Thus, we might expect better survivability of adenine and cytosine over uracil in our haze/sand-mixture experiments, which is indeed the case. It is unclear whether the large sand particles that are typical for the equatorial regions on Titan would produce similar catalytic effects as we see here, given their composition is presently unknown.

It is likely that shocking our haze particle samples is simply breaking biomolecule bonds. However, one alternative possibility worth noting is that shocking these molecules can also lead to the formation of complex macroscale structures. Shock tube experiments by \citet{Singh_et_al2020} found that amino acids tend to form complex agglomerate structures when subjected to 1--3 MPa shocks and 2500--8000 K for 1--2 ms.

Finally, we note two caveats to our experiments: 1) Our experiments were performed without any water in the sample assembly. This is not completely analogous to Titan's icy surface, and we might expect the inclusion of water to modify the biomolecule distributions further (e.g., oxidation). 2) reaction catalysis on the surface of the sample assembly cannot be completely discounted. However, we note that the surface area of the sample assembly is small in comparison to the volume of the sample, or the surface area of the sand grains.

\section*{Conclusions}

In this work, we performed impact experiments on simulated Titan organic haze particles and haze/sand mixtures and measured the biomolecule content using GC/MS/MS.

The four major conclusions are as follows:

\begin{itemize}
\item The nucleobases, amino acids and other biomolecules detected in our simulated Titan haze particles degrade overall when shocked at 13 GPa, by factors of 3--8. However, in some cases certain individual biomolecules, i.e., thymine, xanthine, isoleucine/norleucine, threonine, and aspartic acid are either enriched or even produced from these shocks. This suggests shock synthesis pathways for these species.
\item Uracil, glycine, proline, and tyrosine in haze particles did not survive any impact experiment, suggesting they are the most unstable to these events. This could be because these species react more rapidly than the other species at the temperatures and pressures of our tested shock conditions.
\item Quartz sand may offer catalytic effects for the shock production of threonine and aspartic acid, and the enrichment of xanthine in organic haze particles.
\item Thermal and shock processes affect nucleobase degradation differently. This may be due to the different thermal histories of shock processes and standard heating processes.
\end{itemize}

These experiments are the first attempts to understand the effects of impacts on simulated Titan organic hazes. We intend to expand upon these experiments to further understand and characterize the changes in biomolecule distributions from impacts on Titan's surface. This may involve expanding upon the studied biomolecules, improving upon sterilization procedures, and including water ice in impact experiments.

\section*{Acknowledgements}

We thank the two anonymous referees, whose comments led to improvements in this paper. B.K.D.P. is supported by the NSERC Banting Postdoctoral Fellowship.

\beginsupplement

\section*{Supporting Information}

In Table~\ref{RetentionTimes}, we list the details of the GC/MS/MS gates for the detection of nucleobases, amino acids, and other biomolecules in this study. For each range of retention times (gate), only the parent ion is trapped at the first quadrupole, and, after collision and ion fragmentation at second quadrupole, only the qualifier and quantifier daughter ions are let through the detector. The retention times and collision energies for these species were developed in \citet{Sebree_et_al2018}; not all of these species are detected in the haze particle samples in this study.

\begin{table*}[ht!]
\centering
\caption{Gate retention times, parent ions, and qualifier and quantified daughter ions and their respective collision energies for GC/MS/MS detection of the biomolecules in this study. Gates were developed in \citet{Sebree_et_al2018} and thus not all of these species are detected in the haze particle samples in this work. \label{RetentionTimes}}
\begin{tabular}{lclccc}
\\
\multicolumn{1}{c}{Gate} &  
\multicolumn{1}{l}{Retention time range (min)} & 
\multicolumn{1}{l}{Biomolecule} &
\multicolumn{1}{l}{Parent ion (m/z)} & 
\multicolumn{1}{l}{Daughter ions (m/z)} &
\multicolumn{1}{l}{Collision ener. (eV)}
\\[+2mm] \hline \\[-2mm]
1 & 8.15--9.05 & Ethanolamine & 232 & 147,116 & 10,10\\
2 & 9.05--9.3 & Alanine & 158 & 102,73 & 10,10\\
3 & 9.3--9.7 & Glycine & 218 & 189,147 & 10,10\\
4 & 9.7--9.87 & Sarcosine & 260 & 232,147 & 10,10\\
5 & 9.87--10.23 & $\beta$-Aminoisobutryic Acid & 246 & 147,133 & 10,10\\
6 & 10.23--10.38 & $\beta$-Alanine & 260 & 218,147 & 5,20\\
7 & 10.38--10.8 & Urea & 231 & 173,147 & 15,15\\
 & 10.38--10.8 & Valine/Norvaline & 186 & 130,73 & 12,15\\
8 & 10.8--11.18 & Leucine & 200 & 157,73 & 15,20\\
9 & 11.18--11.58 & Isoleucine/Norleucine & 200 & 157,144,73 & 15,10,20\\
10 & 11.58--12.35 & Proline & 184 & 128,73 & 15,25\\
 & 11.58--12.35 & Uracil & 283 & 169,99 & 30,20\\
11 & 12.35--13.25 & Thymine & 297 & 113,73 & 25,25\\
12 & 13.45--13.87 & Cytosine & 282 & 212,98 & 20,25\\
 & 13.45--13.87 & Methionine & 218 & 170,73 & 10,15\\
13 & 13.87--14.18 & Serine & 288 & 100,73 & 20,20\\
14 & 14.18--14.8 & Threonine & 303 & 202,73 & 20,20\\
15 & 14.8--15.35 & Phenylalanine & 336 & 308,73 & 10,20\\
16 & 15.35--15.65 & Aspartic Acid & 418 & 390,73 & 10,20\\
17 & 15.65--15.87 & Hydroxyproline & 314 & 182,73 & 10,20\\
18 & 15.87--16.04 & Cysteine & 304 & 118,73 & 20,20\\
19 & 16.04--16.4 & Hypoxanthine & 307 & 193,73 & 20,25\\
20 & 16.4--16.73 & Glutamine & 432 & 272,147 & 15,15\\
21 & 16.73--17.15 & Adenine & 306 & 192,73 & 10,25\\
22 & 17.15--18.97 & Lysine & 431 & 300,73 & 10,15\\
23 & 19.15--19.95 & Tyrosine & 466 & 438,147 & 10,15\\
 & 19.15--19.95 & Xanthine & 437 & 363,73 & 20,25\\
24 & 19.95--22.15 & Guanine & 436 & 322,73 & 25,30\\
25 & 22.15--24.35 & Tryptophan & 244 & 188,73 & 10,20\\
\\[-2mm] \hline
\end{tabular}
\end{table*}

\begin{figure*}[!hbtp]
\centering
\includegraphics[width=\linewidth]{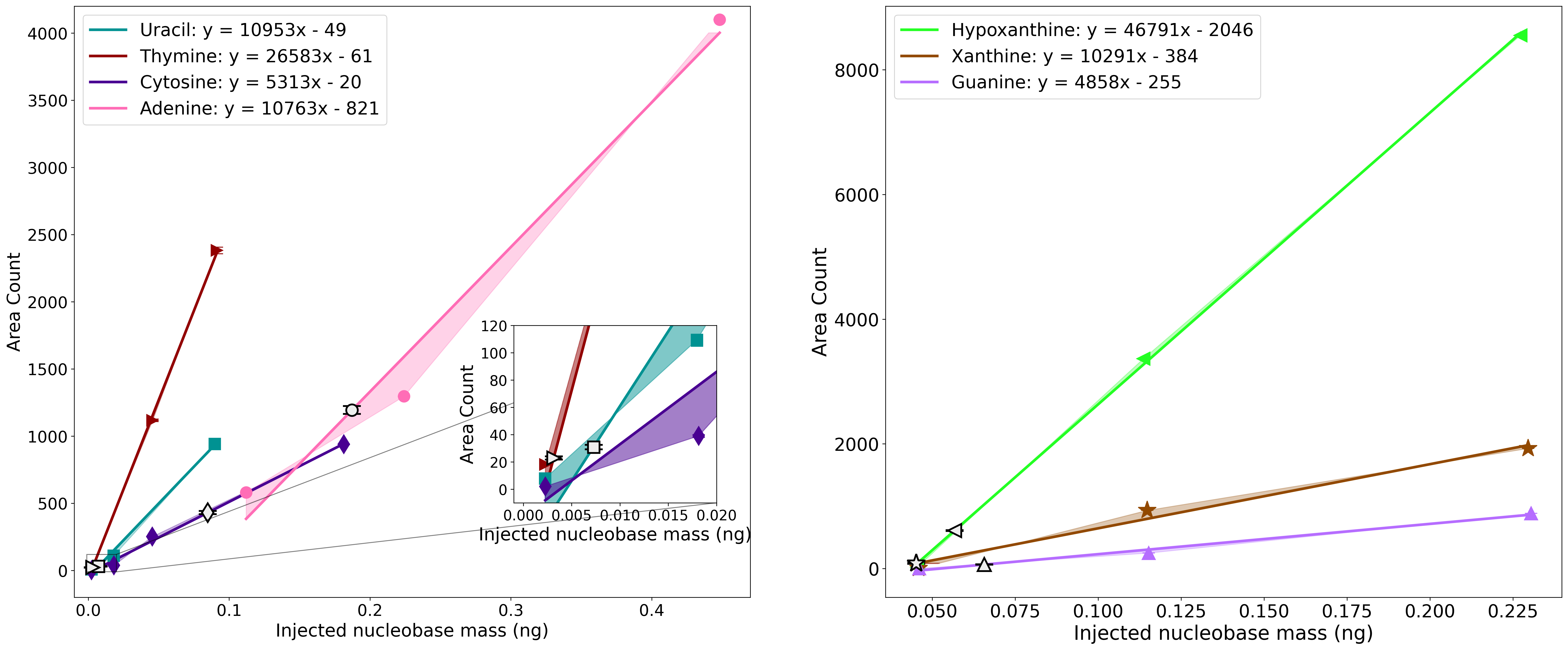}
\caption{Calibration Curves for seven nucleobases. The equations of the lines of best fit are included in the plot legends. Uncertainties are displayed as shaded regions, and are calculated using the variance from a model linearly connecting each data point. As an example calculation, we add the peak areas from the GC/MS/MS analysis of the unshocked haze particles. \label{NucleobaseCalCurve}}
\end{figure*}

\begin{figure*}[!hbtp]
\centering
\includegraphics[width=\linewidth]{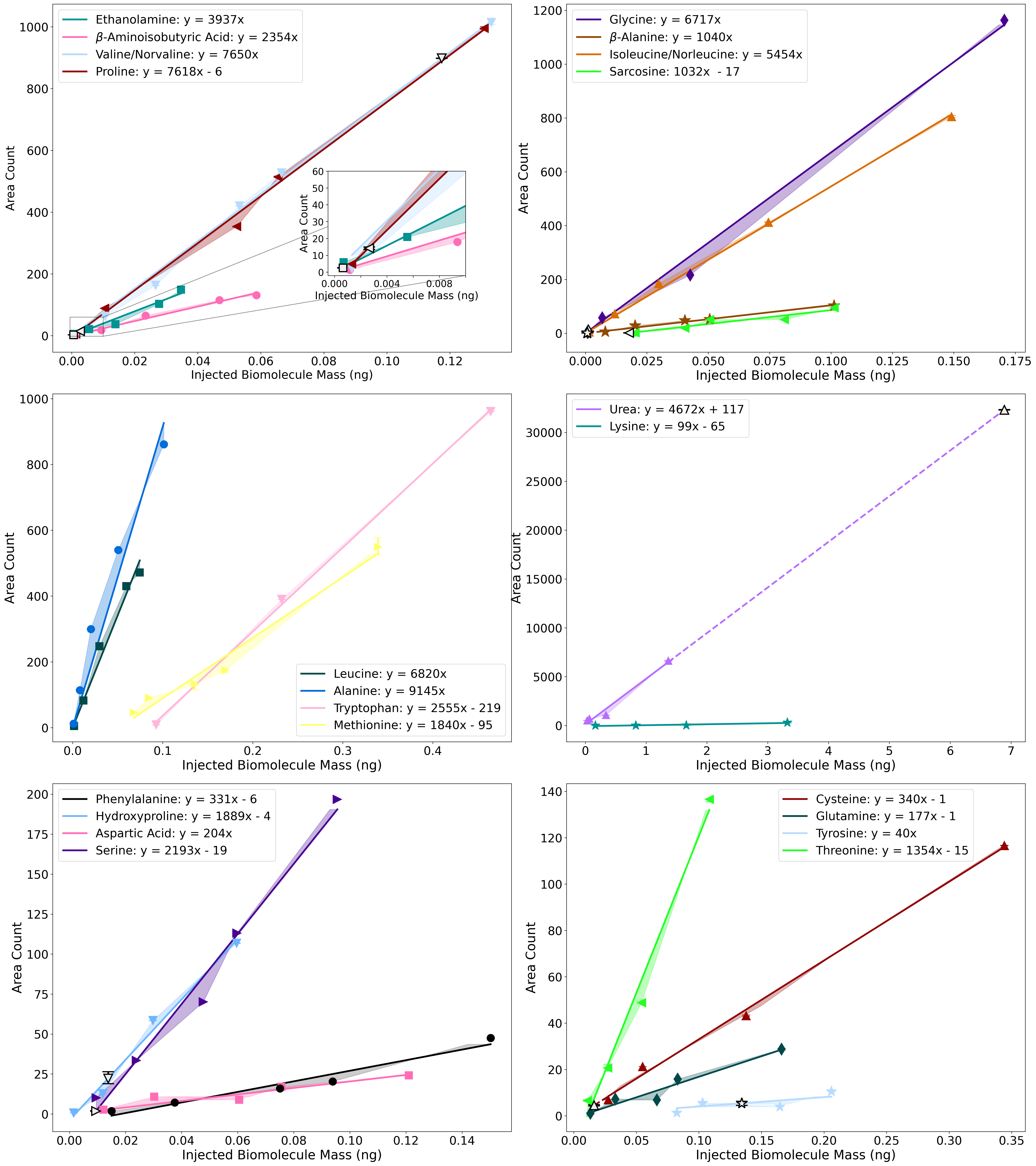}
\caption{Calibration Curves for 22 amino acids and other biomolecules. The equations of the lines of best fit are included in the plot legends. Uncertainties are displayed as shaded regions, and are calculated using the variance from a model linearly connecting each data point. As an example calculation, we add the peak areas from the GC/MS/MS analysis of the unshocked haze particles. \label{PhysioCalCurve}}
\end{figure*}

In Figures~\ref{NucleobaseCalCurve} and \ref{PhysioCalCurve}, we display the calibration curves for the quantification of nucleobases and amino acids/other biomolecules, respectively. 

In Table~\ref{LODLOQTable}, we list the biomolecule injection masses calculated for all of our samples based on their GC peak areas, and compare them to the limits of detection (LOD) and limits of quantification (LOQ) roughly constrained by our calibration curve data. The LOD is defined as the lowest injection mass capable of producing a signal with a SNR$\sim$3, and the LOQ is defined as the lowest injection mass capable of producing a singal with a SNR$\sim$10. Given most of calibration curve data has a SNR$>$10, we can often only provide upper bounds on the LOQ. When a biomolecule peak falls below the rough upper bound LOQ or LOD, we also display the SNR for that peak.

All of the sample nucleobase detections are above the upper bound LOQs. However, two proteinogenic amino acid detections and a few other biomolecule detections are below the upper bound LOQs. The aspartic acid detection has a SNR of 21, suggesting this detection is above the true LOQ. The isoleucine/norleucine, and cysteine detections have SNRs of 8 and 3.3, respectively; therefore, we display these detections as upper limits in Table~\ref{ConcentrationTable}. Similarly, the $\beta$-alanine and sarcosine detections, and two of the ethanolamine detections have 3 $<$ SNR $<$10; thus, we also treat these detections as upper limits.

\begin{table*}[ht!]
\centering
\caption{Injection masses measured for each sample in comparison to limits of detection (LOD), i.e., SNR$\sim$3, and limits of quantification (LOQ), i.e, SNR$\sim$10. The LOD and LOQ are roughly constrained by the lower calibration curve data. Often, only upper bound LOD or LOQ are possible, as the SNR for the lowest standard concentration was often $>$ 3 or $>$ 10, respectively. In the cases where the injection masses for our samples are lower than the upper bound LOD or LOQ, we display the SNR for that measurement. \label{LODLOQTable}} 
\resizebox{\textwidth}{!}{\begin{tabular}{lcccccccc}
\\
\multicolumn{1}{l}{Biomolecule} &  
\multicolumn{1}{l}{Unshock. Haze} & 
\multicolumn{1}{l}{Shock. Haze} &
\multicolumn{1}{l}{Shock. H:S 50:50} & 
\multicolumn{1}{l}{Shock. H:S 10:90} &  
\multicolumn{1}{l}{Unshock. Sand} &  
\multicolumn{1}{l}{Shock. Sand} &  
\multicolumn{1}{l}{LOD} & 
\multicolumn{1}{l}{LOQ}
\\[+2mm] \hline \\[-2mm]
{\bf Nucleobases} & &  \\
Adenine & 187 & 16 & 10 & 12 & - & - & $<$2 & $<$2 \\
Cytosine & 85 & 6 & 6 & 6 & - & - & $<$2 & $<$2 \\
Guanine & 66 & 41 & 25 & 45 & - & - & $<$2 & $\sim$18 \\
Hypoxanthine & 57 & 58 & 30 & 75 & 37 & 23 & $<$18 & $\sim$18 \\
Xanthine & 45 & 24 & - & 113 & 52 & 25 & $<$2 & $<$2 \\
Uracil & 7 & - & - & - & - & - & $<$2 & $<$2 \\
Thymine & 3 & 6 & 3 & - & 2.3 & - & $<$2 & $<$2 \\
{\bf Proteinogenic Amino Acids} & & \\
Tyrosine & 134 & - & - & - & - & - & $<$82 & $<$82 \\
Valine/Norvaline & 117 & 53 & 70 & 58 & - & - & $<$1.3 & $<$11\\
Cysteine & 16 (SNR=3.3) & - & - & - & - & - & $<$28 & $<$28 \\
Proline & 3 & - & - & - & - & - & $<$1.3 & $<$1.3 \\
Glycine & 1.1 & - & - & - & - & - & $<$0.85 & $<$0.85 \\
Isoleucine/Norleucine & - & 0.8 (SNR=8) & - & - & - & - & $<$1.5 & $<$1.5 \\
Aspartic Acid & - & - & 12 (SNR = 21) & - & - & - & $\sim$12 & $<$30 \\
Threonine & - & - & 11 & - & - & - & $<$11 & $\sim$11 \\
Serine & 10 & - & 11 & - & - & - & $<$10 & $<$10 \\
Alanine & - &  - & - & - & - & - & $<$1 & $<$1 \\
Leucine & - - & - & - & - & - & - & $<$1.5 & $<$1.5 \\
Hydroxyproline & - & - & - & - & - & - & $<$1.5 & $<$12\\
Phenylalanine & - & - & - & - & - & - & $<$15 & $<$15 \\
Glutamine & - & - & - & - & - & - & $<$13 & $<$13 \\
Methionine & - & - & - & - & - & - & $<$68 & $<$68 \\
Cysteine & - & - & - & - & - & - & $<$28 & $<$28 \\
Tyrosine & - & - & - & - & - & - & $<$82 & $<$82 \\
Lysine & - & - & - & - & - & - & $<$166 & $<$166 \\
{\bf Other Biomolecules} & & \\
Urea & 6891 & 1706 & 1422 & 2167 & 12.5 (SNR=41) & - & $<$14 & $<$34 \\
$\beta$-Alanine & 0.8 (SNR=3.2) & - & - & - & - & - & $<$8 & $<$8 \\
Sarcosine & 18 (SNR=4) & - & - & - & - & - & $<$20 & $<$20 \\
Ethanolamine & 0.65 (SNR=11) & 0.4 (SNR=11) & 0.55 (SNR=42) & 0.13 (SNR=8) & 0.55 (SNR=11) & 0.43 (SNR=9.4) & $<$0.7 & $<$0.7 \\
$\beta$-Aminoisobutyric Acid & - & - & - & - & - & - & $<$1.2 & $<$9 \\
\\[-2mm] \hline
\end{tabular}}
\end{table*}

\bibliography{Bibliography}
\bibliographystyle{aasjournal}

\end{document}